\documentclass[aps, 10pt, amsmath, twocolumn, superscriptaddress, longbibliography]{revtex4-1}
\usepackage[plainpages=false,colorlinks=true,linkcolor=blue, citecolor=blue, urlcolor=blue]{hyperref}
\usepackage[english]{babel}
\usepackage{graphicx}
\usepackage[charter]{mathdesign}
\usepackage{bm}
\usepackage[svgnames]{xcolor}
\usepackage{subfigure}
\usepackage{booktabs}

\usepackage{array}

\newcolumntype{C}{>{$\displaystyle}c<{$}}
\newcolumntype{L}{>{$\displaystyle}l<{$}}
\newcolumntype{R}{>{$\displaystyle}r<{$}}
\newcolumntype{A}{ >{$} r <{$} @{\extracolsep{0pt}} >{${}} l <{$} }
\begin{document}
\title{Superconductivity in the twisted bilayer transition metal dichalcogenide WSe${}_2$ : a quantum cluster study}

\author{Mathieu B\'elanger}
\affiliation{D\'epartement de physique and Institut quantique, Universit\'e de Sherbrooke, Sherbrooke, Qu\'ebec, Canada J1K 2R1}
\author{J\'er\^ome Fournier}
\affiliation{D\'epartement de physique and Institut quantique, Universit\'e de Sherbrooke, Sherbrooke, Qu\'ebec, Canada J1K 2R1}
\author{David S\'en\'echal}
\affiliation{D\'epartement de physique and Institut quantique, Universit\'e de Sherbrooke, Sherbrooke, Qu\'ebec, Canada J1K 2R1} 
\date{\today}

\begin{abstract}

The observation of flat energy bands in transition metal dichalcogenide bilayers such as twisted WSe${}_2$ makes those materials interesting prospects for reproducing the behavior observed in graphene-based systems. We use an effective Hubbard model providing a description of twisted WSe${}_2$ to explore the presence of superconductivity, which was previously reported in experiments. Using both the variational cluster approximation and cluster dynamical mean-field theory, we predict the existence of chiral supercondictivity of type $d\pm id$ that can be tuned by the twist angle and by the application of a perpendicular displacement field, in both electron- and hole-doped systems.
\end{abstract}
%\pacs{}
\maketitle

%===============================================================================
\section{Introduction}

The observation of correlated insulator behavior and unconventional superconductivity in twisted bilayer graphene \cite{cao_correlated_2018,cao_unconventional_2018} has stimulated research on twisted van der Waals heterostructures. Those features appear close to a ``magic'' twist angle, $\theta\approx 1.1^\circ$, that had been predicted by previous theoretical work \cite{trambly_de_laissardiere_localization_2010,bistritzer_moire_2011}. Similar features have been observed in other graphene-based twisted systems \cite{park_tunable_2021,hao_electric_2021,cao_pauli-limit_2021}.

One of the issues with twisted bilayer graphene is the difficulty of reproducing the observations. A small deviation of the twist angle results in a large variation of the critical temperature, critical magnetic field, critical current and density range \cite{balents_superconductivity_2020}. This can be explained by the semi-metallic nature of graphene. Correlated features arise from the flat energy bands. Such bands can be observed near the $K$ valley of graphene, but only close to the magic angle. A small deviation from this magic angle has a big effect on the flatness of the bands, and then this leads to experimental features that are difficult to reproduce exactly.  

Similar correlated states have been observed and predicted in other van der Waals systems that present a moir\'e pattern, like twisted bilayer boron nitride \cite{xian_multiflat_2019} and transition metal dichalcogenides (TMD). The latter can occur in a homobilayer ($MX{}_2/MX{}_2$) \cite{wang_correlated_2020,an_interaction_2020,zhang_flat_2020,venkateswarlu_electronic_2020,ghiotto_quantum_2021} or heterobilayer ($MX_2/M'X'_2$) \cite{phillips_commensurate_2019,ruiz-tijerina_interlayer_2019,regan_mott_2020,tang_simulation_2020,brotons-gisbert_moire-trapped_2021,huang_correlated_2021,li_imaging_2021,morales-duran_metal-insulator_2021,vitale_flat_2021}, where $M$ stands for the metal and $X$ for the chalcogen. One of the advantages of TMD is the existence of a continuum of magic angles with correlated insulating states. The semiconducting nature of TMD leads to a band structure from the monolayer that is flatter than that of graphene. It is then easier to observe flat bands for a larger spectrum of angles. This makes correlated behavior easier to observe and reproduce. \cite{balents_superconductivity_2020}. Another interesting feature of TMD is the valley-dependent spin splitting caused by spin-orbit coupling \cite{xiao_coupled_2012}. Depending on the strength of the spin-orbit splitting at the Brillouin zone corners, the valence band maximum can be at the $\pm K$ or $\Gamma$ point \cite{angeli__2021}. One TMD of particular interest is twisted WSe${}_2$, where correlated insulating states and evidence of superconductivity have been observed \cite{wang_correlated_2020,an_interaction_2020,zhang_flat_2020}.

A strategy has been proposed to obtain effective, low-energy models of twisted TMD  \cite{jung_ab_2014}. It can be applied to specific materials like MoTe${}_2$ and WSe${}_2$ \cite{wu_topological_2019,pan_band_2020}.
Following this approach, twisted WSe${}_2$ can be described by a two-dimensional Hubbard model on a triangular lattice with complex-valued hopping amplitudes to the third nearest-neighbor. Moreover, the effect of a voltage bias across the bilayer is to change the relative phases of spin-up and spin-down hopping amplitudes \cite{pan_band_2020}.

A $120^\circ$ magnetic order is expected at half-filling in the simple, nearest-neighbor Hubbard model on a triangular lattice \cite{weber_magnetism_2006,laubach_phase_2015,misumi_mott_2017,wietek_mott_2021}. Away from half-filling, superconductivity of type $d+id$ has been predicted to exist \cite{weber_magnetism_2006,chen_unconventional_2013}. The effective moir\'e-band Hubbard model, although also based on the triangular lattice, has a very different dispersion relation from the simple nearest-neighbour model, but similar magnetic phases are predicted  \cite{pan_band_2020,zang_hartree-fock_2021,zang_dynamical_2022}. Chiral superconductivity has been predicted using renormalization group analysis in the weak coupling regime for twisted bilayers TMD with only nearest-neighbour hopping \cite{wu_pair-density-wave_2022}.

In this paper, we use quantum cluster methods, namely the variational cluster approximation (VCA) and the cluster dynamical mean-field theory (CDMFT), to argue that superconductivity of type $d+id$ occurs in the effective Hubbard model describing twisted WSe${}_2$ with large values of interaction and hopping up to the third nearest neighbor. The superconducting domes can be tuned by modifying the twist angle and by using an external displacement field. The complex hopping terms introduced by the displacement field are a major factor in determining the shape and location of the superconducting dome. We did not take into consideration the $120^\circ$ antiferromagnetic magnetic phase predicted in Ref \cite{pan_band_2020}, since no competition with the superconducting phase is expected in the doping region considered (the magnetic phase is concentrated close to half-filling).

This paper is organized as follow. In Sect. \ref{sec:model}, we introduce the moir\'e Hubbard model. Then in Sect. \ref{sec:numerical} we present the two numerical methods used : VCA and CDMFT. In Sects. \ref{sec:electron} and \ref{sec:hole}, we present our results for the electron- and hole-doped sides, respectively. We discuss some implications of our results in section \ref{sec:discussion}.

%===============================================================================
\section{Model}
\label{sec:model}
The electronic states in twisted layered materials like bilayer graphene and hBN can be described by a moir\'e band model \cite{jung_ab_2014}, from which one can extract a number of bands equal to the number of atoms present in the moir\'e unit cell. The same ideas can be applied to TMD homobilayers like MoTe${}_2$ \cite{wu_topological_2019} and WSe${}_2$ \cite{pan_band_2020} with strong spin-orbit splitting at the $\pm K$ point of the Brillouin zone. A moir\'e model includes the effect of the interlayer tunneling between two layers of similar or identical lattice structures. A layer-dependent moir\'e potential describes the effect of the twist angle $\theta$ between the two layers. One can then tune this moir\'e potential and interlayer tunneling to reproduce the density of states from STM experiments \cite{zhang_flat_2020}. This also reproduces the local density of states around the high-symmetry positions where the atoms of the two layers sit on top of each other. It is also possible to introduce the effect of an external perpendicular displacement field that affects the moir\'e potential. From this moir\'e band model, one can construct an effective low-energy tight-binding model to study a few moir\'e valence bands. 

To study the possibility of superconductivity in twisted WSe${}_2$, we use the model proposed in Ref. \cite{pan_band_2020}. In that work, the authors constructed an effective Hubbard model for the first moir\'e valence band. To do so, they extracted the moir\'e Wannier function from the moir\'e band model \cite{wu_topological_2019}. This method has been used to describe various homobilayer and heterobilayer TMDs \cite{wu_topological_2019,pan_band_2020}.

The Hamiltonian of this effective Hubbard model is expressed as
\begin{align}
H=\sum_{\sigma}\sum_{\mathbf{r},\mathbf{r'}}t_{\mathbf{r}\mathbf{r'},\sigma }c^\dagger_{\mathbf{r},\sigma}c_{\mathbf{r'},\sigma}+U\sum_{\mathbf{r}}n_{\mathbf{r},\uparrow}n_{\mathbf{r},\downarrow},
\label{eq:H}
\end{align}
where $\mathbf{r},\mathbf{r'}$ are sites on the triangular lattice and $\sigma=\uparrow,\downarrow$ represents the spins associated with the $+K$ and $-K$ valleys, respectively. $c_{\mathbf{r},\sigma}$ ($c^\dagger_{\mathbf{r},\sigma}$) destroys (creates) an electron at site $\mathbf{r}$ with spin $\sigma$. $n_{\mathbf{r},\sigma}=c^\dagger_{\mathbf{r},\sigma}c_{\mathbf{r},\sigma}$ is the number of electrons on site $\mathbf{r}$ with spin $\sigma$. $U$ is the on-site repulsion between electrons. The hopping matrix $t_{\mathbf{r}\mathbf{r'},\sigma}$ is Hermitian : $t_{\mathbf{r}\mathbf{r'},\sigma}=t^*_{\mathbf{r'}\mathbf{r},\sigma}$. The system has a threefold rotation symmetry and three reflections across mirror planes perpendicular to the $xy$ plane, corresponding to the $C_{3v}$ symmetry group. Finally, time-reversal symmetry $(\mathcal{T})$ requires that $t_{\mathbf{r}\mathbf{r'},\sigma}=t^*_{\mathbf{r}\mathbf{r'},-\sigma}$. Considering those relations, it is possible to express the hopping parameters between the $n$-th nearest neighbors as $t_{\sigma n}=|t_n|e^{i\phi_n^\sigma}$, where the amplitude $|t_n|$ and the phase $\phi_n^\sigma$ depend on the twist angle $\theta$ and on the inter layer potential $V_z$. The on-site interaction $U$ also depends on $\theta$, $V_z$ and $\epsilon$, the effective background dielectric constant, which can change depending on the experimental substrate. The value of the parameters considered here are extracted from Ref. \cite{pan_band_2020}. The hopping parameters grow quickly with the twist angle $\theta$. This can be explained by the Wannier states getting physically closer as the moir\'e period shortens, as mentionned in Ref [28]. The norms of the hopping parameters also grow with the external displacement field, but seem to reach a plateau for high values of $V_z$ (not considered in the present paper). The value of the on-site interaction grows as a function of the twist angle, because of the smaller size of the Wannier function, but more slowly than the hopping parameters. Since every parameter in the present work is normalized by $t_1$, the resulting effect is that the value of $U/t_1$ decrease with twist angle. $V_z$ do not affect the on-site interaction, resulting in a decrease of $U/t_1$ with $V_z$.

The twist angle is defined in relation to an axis going through a AA stacking site where metal or chalcogen atoms are perfectly aligned with the same type of atom in the other layer. We consider hopping up to the third ($n=3$) nearest neighbor. We neglect interactions between different sites. Every parameter is expressed in units of $|t_1|$. We limit ourselves to twist angles $\theta\ge3^\circ$ since lattice relaxation effects, that are not taken into account in this model, become important for small twist angles ($\theta<2.5^\circ$) \cite{pan_band_2020,enaldiev_stacking_2020}.

When only nearest-neighbor hopping is present ($t_{\sigma2}=t_{\sigma3}=0$), there is a particle-hole symmetry under $n\rightarrow2-n$ and $\phi\rightarrow \pi-\phi$. This symmetry is lost with the addition of next-nearest-neighbor hopping. We then expect different properties on the electron- and hole-doped sides. This explains differences with some of the previous work on the triangular lattice Hubbard model \cite{wu_pair-density-wave_2022}.

Model (\ref{eq:H}) can be considered as a triangular Hubbard model where control over the different parameters is obtained by tuning $\theta$, $V_z$ and $\epsilon$. The simple nearest-neighbor triangular lattice Hubbard model has been studied previously with prediction of a $120^\circ$ antiferromagnetic order at half-filling \cite{weber_magnetism_2006,laubach_phase_2015,misumi_mott_2017,wietek_mott_2021}. This order as been predicted in model (\ref{eq:H}) for a range of $V_z$ \cite{pan_band_2020,zang_hartree-fock_2021,zang_dynamical_2022}. The simple nearest-neighbor triangular lattice Hubbard is also predicted to have a type $d+id$ superconductivity phase when doped away of half-filling \cite{weber_magnetism_2006,chen_unconventional_2013}. Chiral superconductivity, including type $d+id$, has been predicted using renormalization group analysis in the weak interaction limit for twisted bilayers TMD \cite{wu_pair-density-wave_2022}. We do expect to observe a similar phase in the twisted WSe${}_2$ in the strongly interacting case, modulated by the different control parameters.

The presence of complex hopping terms up to the third nearest neighbor in model (\ref{eq:H}) distinguishes our work from previous studies on the simple, nearest-neighbor triangular lattice Hubbard model. Another difference from previous work on superconductivity in twisted WSe${}_2$ is our ability to treat large values of the interaction $U$. 

%~~~~~~~~~~~~~~~~~~~~~~~~~~~~~~~~~~~~~~~~~~~~~~~~~~~~~~~~~~~~~~~~~~~~~~~~~~~~~~~
\begin{figure}[h]
	\includegraphics[scale=0.75]{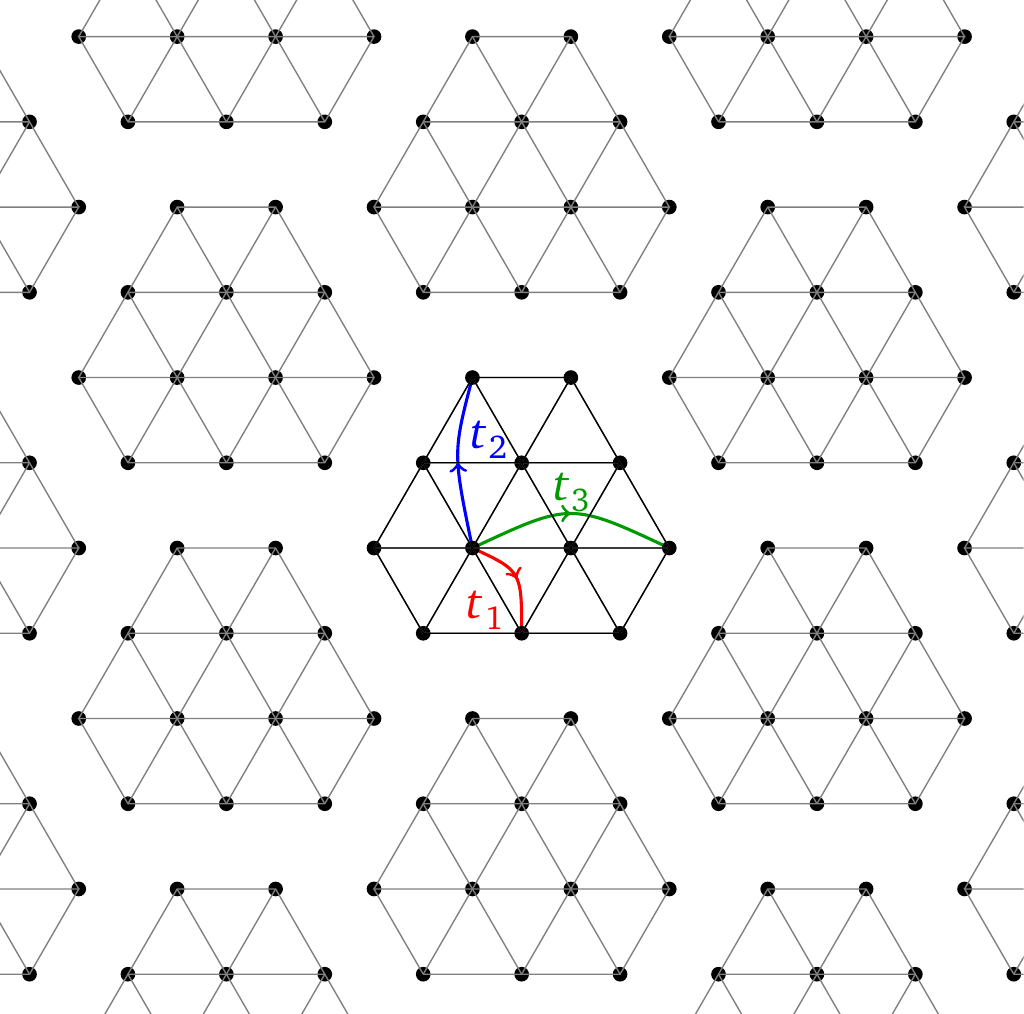}
	\caption{Cluster used in the VCA method. The different hopping parameters of model (\ref{eq:H}) are illustrated.
	}
	\label{fig:schema_model}
\end{figure}
%~~~~~~~~~~~~~~~~~~~~~~~~~~~~~~~~~~~~~~~~~~~~~~~~~~~~~~~~~~~~~~~~~~~~~~~~~~~~~~~

\section{Numerical methods}
\label{sec:numerical}

In order to probe the possibility of superconductivity in model (\ref{eq:H}), we use quantum cluster methods with an exact diagonalization solver at zero temperature. Specifically, we use two of those methods: the variational cluster approximation (VCA) and cluster dynamical mean-field theory (CDMFT). 

In those methods, the lattice is tiled into an infinite number of identical clusters. Two systems are then considered : the original system on the infinite lattice, described by the Hamiltonian $H$, and the ``reference'' system defined only on the cluster, described by the Hamiltonian $H'$. The only requirement is that both systems $H$ and $H'$ share the same interaction term. 

Probing the various broken symmetries is done using an embedding scheme on $H'$. In the VCA, $H'$ is the restriction of $H$ on the cluster, augmented by one or more {\it Weiss fields} representing the broken symmetry operators being probed. In CDMFT, the effect of the environment of the cluster is added to $H'$ via a set of uncorrelated orbitals (the {\it bath}). In both cases, an optimal self-energy can be extracted from $H'$ and applied to $H$. In other words, the approximate Green function $\mathbf{G}$ of the original lattice Hamiltonian is constructed by inserting the self-energy $\mathbf{\Sigma}$ found from $H'$ in Dyson's equation:
\begin{align}
\mathbf{G}(\mathbf{k},\omega)=\frac{1}{\mathbf{G}_0^{-1}(\mathbf{k},\omega)-\mathbf{\Sigma}(\omega)}.
\label{eq:Dyson}
\end{align}
In this expression the wavevectors $\mathbf{k}$ are restricted to the reduced Brillouin zone (rBZ) of the superlattice of clusters. $\mathbf{G}_0$ is the non-interacting Green function of the infinite lattice. $\mathbf{G}$, $\mathbf{G}_0$ and $\mathbf{\Sigma}$ are $2L\times2L$ matrices, $L$ being the number of sites on the cluster (the factor of 2 because of spins). From the Green function, we can compute the average of any one-body operator, like the order parameter associated with superconductivity. 

To explain the last point, let us consider a matrix $s_{\alpha\beta}$ that defines the one-body operator $\hat{S}$
\begin{align}
\hat{S}=\sum_{\alpha\beta,\mathbf{k}}s_{\alpha\beta}(\mathbf{k})c_\alpha^\dagger(\mathbf{k}) c_\beta(\mathbf{k}).
\end{align}
The indices $\alpha,\beta$ stand for a composite of cluster site and spin indices. The expectation value of $\hat{S}$ is
\begin{align}
\langle \hat{S}\rangle =\int\frac{d\omega}{2\pi}\int_{\text{rBZ}}\frac{d^2k}{(2\pi)^2}\text{tr}[\mathbf{s}(\mathbf{k})\mathbf{G}(\omega,\mathbf{k})]
\label{eq:expectation}
\end{align} 
where the frequency integral is taken over a contour that circles the negative real axis, targeting only the occupied states.

The size of each cluster should allow one to compute the electron Green function numerically. The exact diagonalization method used to compute the cluster Green function limits the total number of orbitals used. The VCA procedure allows one to consider a larger cluster than CDMFT, because of the baths orbitals that need to be added in the latter (see section \ref{sec:cdmft} for details). This leads to a better account of spatial correlations in VCA, whereas CDMFT has a better treatment of temporal correlations. Both methods allow us to go beyond mean-field theory while keeping the fully correlated character of the model within each cluster. It is possible to apply those methods to strongly correlated systems to probe various broken symmetry phases \cite{dahnken_variational_2004,sahebsara_hubbard_2008,faye_interplay_2017,kancharla_anomalous_2008,dash_pseudogap_2019}. 

Let us mention that the geometric frustration of the triangular lattice and the complex hopping amplitudes enhance the sign problem in Monte Carlo methods. This further justifies the use of the exact diagonalization solver in this work.

\subsection{The Variational Cluster Approximation}

The VCA is a variational method on the electron self-energy based on Potthoff's self-energy functional approach \cite{potthoff_variational_2003,dahnken_variational_2004}. This method as been used to predict magnetic phases \cite{dahnken_variational_2004,sahebsara_hubbard_2008} and superconductivity \cite{faye_interplay_2017}. For a detailed review of the method, see Refs \cite{potthoff_variational_2012,potthoff_cluster_2018}.

To determine the optimal one-body part of $H'$ in VCA, the electron self-energy $\mathbf{\Sigma}$ associated with $H'$ is used as a variational self-energy to construct the Potthoff self-energy functional \cite{dahnken_variational_2004}:

\begin{align}
\begin{split}
\Omega[\mathbf{\Sigma}(\xi)]=&\Omega'[\mathbf{\Sigma}(\xi)]+\text{Tr}\ln[-(\mathbf{G}_0^{-1}-\mathbf{\Sigma}(\xi))^{-1}]\\
&-\text{Tr}\ln(-\mathbf{G}'(\xi)),
\end{split}
\label{eq:potthoff}
\end{align}
with $\mathbf{G}'$ the physical Green function of the cluster. The symbol $\xi$ stands for a collection of parameters that define the one-body part of $H'$. Tr stand for a functional trace, which implies a sum over frequencies, momenta and bands. Finally, $\Omega'$ is the exact grand potential of $H'$, i.e., the ground state energy if the chemical potential $\mu$ is included in the Hamiltonian. We use the Lanczos method at zero temperature to compute numerically $\mathbf{G}'(\omega)$ and $\Omega'$. The Potthoff functional $\Omega[\mathbf{\Sigma}(\xi)]$ is computed exactly but only on a restricted space of $\mathbf{\Sigma}(\xi)$ that are the physical self-energy of the reference system $H'$. 

The optimal value of the self-energy $\mathbf{\Sigma}$ corresponds to stationary value of $\Omega(\xi)$. We use a standard optimization method (Newton-Raphson) in the space of the one-body parameter $\xi$ to solve 
\begin{align}
\frac{\partial\Omega(\xi)}{\partial \xi}=0.
\label{eq:min}
\end{align}
We then construct the approximate Green function $\mathbf{G}$ of the original lattice Hamiltonian by inserting the self-energy obtained here in Eq. (\ref{eq:Dyson}), and extract the order parameter using Eq. (\ref{eq:expectation}).

To study the superconducting phase in twisted WSe${}_2$, we use a 12-site cluster that respects the $C_{3v}$ symmetry of model (\ref{eq:H}) as shown in Fig. \ref{fig:schema_model}. We expect the superconducting order parameter to fall into one of the irreducible representations (irrep) of $C_{3v}$. Clusters that do not respect this symmetry do not allow us to probe the different irreps separately. Finally, the exact diagonalization used to find the Green function $\mathbf{G}'$ of the cluster  limits the maximum number of sites we can use. All those reasons justify the use of the 12-site cluster shown in Fig \ref{fig:schema_model}.

The Weiss fields that define the set of one-body parameters $\xi$ are chosen to correspond to one of the irrep of $C_{3v}$. We first define the pairing operators between two adjacent sites as
\begin{align}
\text{singlet:}\quad S_{\mathbf{r},i}=c_{\mathbf{r},\uparrow}c_{\mathbf{r}+\mathbf{e}_i,\downarrow}-c_{\mathbf{r},\downarrow}c_{\mathbf{r}+\mathbf{e}_i,\uparrow}\\
\text{triplet:}\quad T_{\mathbf{r},i}=c_{\mathbf{r},\uparrow}c_{\mathbf{r}+\mathbf{e}_i,\downarrow}+c_{\mathbf{r},\downarrow}c_{\mathbf{r}+\mathbf{e}_i,\uparrow}
\end{align}
We then define pairing operators falling in the different irreps of $C_{3v}$ by the combinations given in Table \ref{table:pairing}. We note that the chiral pairings $d+id$ and $d-id$ are degenerate and so a VCA solution found with one of them would also exist with the other (same for $p+ip$ and $p-ip$).

\begin{table}[h]

	\begin{ruledtabular}
		\begin{tabular}{ccA}
			Irrep&Symbol&\multicolumn{2}{c}{Operators}\\
			\hline
			$A_1$&$s$&\hat{\Delta}_s&=\sum_{\mathbf{r}}\left(S_{\mathbf{r},1}+S_{\mathbf{r},2}+S_{\mathbf{r},3}\right)\\
			
			$A_2$&$f$&\hat{\Delta}_f&=\sum_{\mathbf{r}}\left(T_{\mathbf{r},1}+T_{\mathbf{r},2}+T_{\mathbf{r},3}\right)\\
			
			$E_1$&$d+id$&\hat{\Delta}_{d+id}&=\sum_{\mathbf{r}}\left(S_{\mathbf{r},1}+\omega S_{\mathbf{r},2}+\bar{\omega}S_{\mathbf{r},3}\right)\\	
			&$d-id$&\hat{\Delta}_{d-id}&=\sum_{\mathbf{r}}\left(S_{\mathbf{r},1}+\bar{\omega}S_{\mathbf{r},2}+\omega S_{\mathbf{r},3}\right)\\
			&$p+ip$&\hat{\Delta}_{p+ip}&=\sum_{\mathbf{r}}\left(T_{\mathbf{r},1}+\omega T_{\mathbf{r},2}+\bar{\omega}T_{\mathbf{r},3}\right)\\	
			&$p-ip$&\hat{\Delta}_{p-ip}&=\sum_{\mathbf{r}}\left(T_{\mathbf{r},1}+\bar{\omega}T_{\mathbf{r},2}+\omega T_{\mathbf{r},3}\right)	
		\end{tabular}
	\end{ruledtabular}
	\caption{Pairing operators used as Weiss field in the VCA procedure with $\omega=e^{\frac{2\pi i}{3}}$ and $\bar{\omega}=e^{-\frac{2\pi i}{3}}$. Each belongs to an irreducible representation (irrep) of the $C_{3v}$ point group. Chiral pairings belong to the $E_1$ irrep.\label{table:pairing}}
\end{table}

\subsection{Cluster dynamical mean-field theory}
\label{sec:cdmft}
In CDMFT, each cluster in $H'$ is augmented by an adjustable environment made of a set of uncorrelated orbitals (the {\it bath}). Each cluster then defines an Anderson impurity model that needs to be solved by an impurity solver to extract the Green function of the cluster $\mathbf{G}_c(\omega)$. Like before, we use exact diagonalization as our impurity solver at zero temperature. By using Dyson's equation we extract the cluster self-energy $\mathbf{\Sigma(\omega)}$
\begin{align}
\mathbf{G}_c^{-1}(\omega)=\omega-\mathbf{t}_c-\mathbf{\Gamma}(\omega)-\mathbf{\Sigma}(\omega)
\end{align}
where $\mathbf{t}_c$ is the hopping matrix on the cluster and $\mathbf{\Gamma}(\omega)$ is the hybridization function that contains the information about the uncorrelated bath orbitals. We then use the cluster self-energy as an approximation of the full lattice Green function in Eq. (\ref{eq:Dyson}). The bath parameters are chosen to minimize the difference between  $\mathbf{G}_c(\omega)$ and the local version of $\mathbf{G}(\mathbf{k},\omega)$. We use a self-consistency procedure to do so. More details on this method can be found in Refs \cite{kancharla_anomalous_2008,senechal_bath_2010,senechal_quantum_2015}.

\begin{figure}[t]
	\begin{center}
	\includegraphics[scale=0.5]{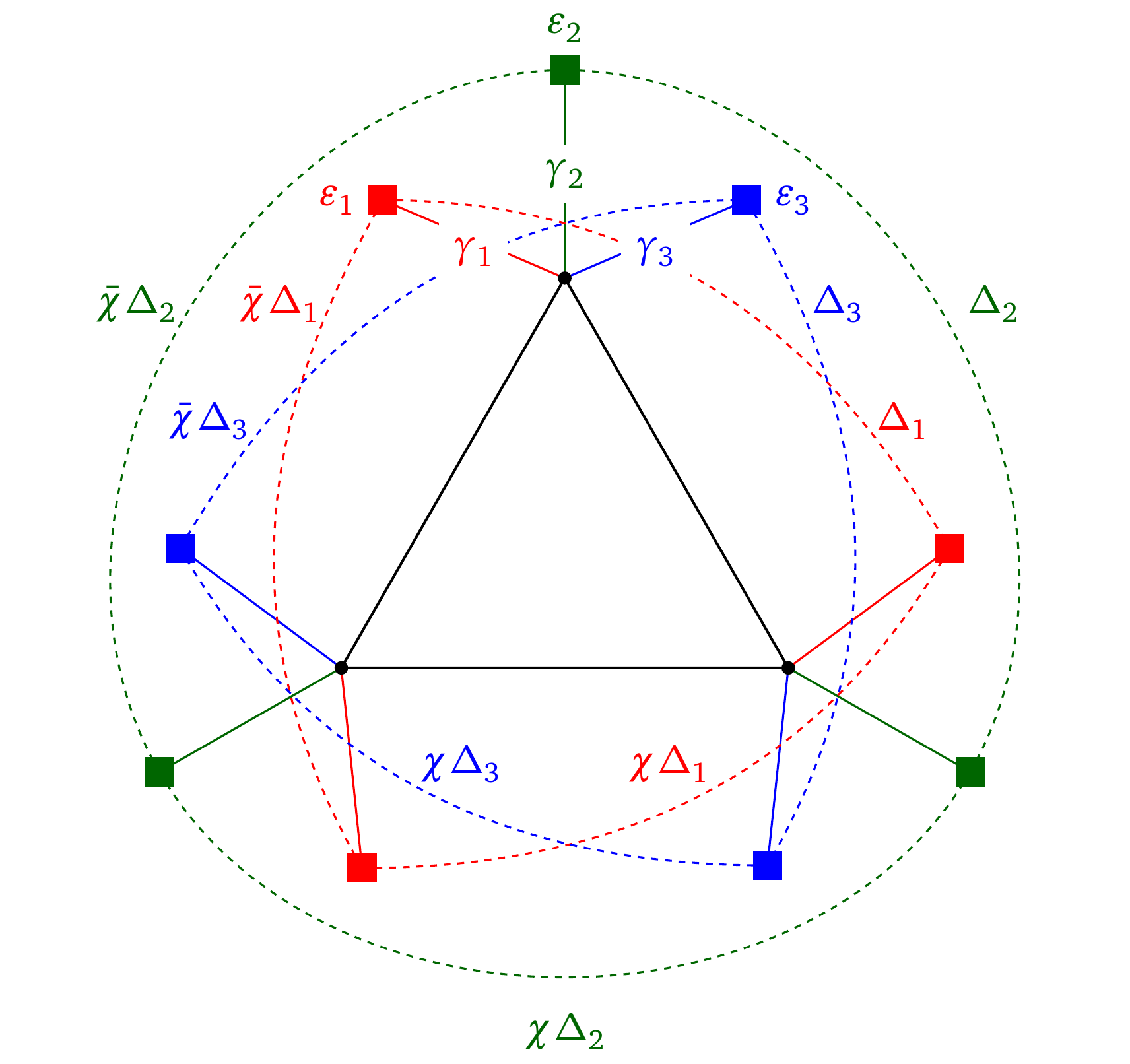}
	\caption{Impurity model used in the CDMFT procedure with a 3-site cluster and 9 uncorrelated bath orbitals. The latter are separated in three groups (red, green, blue) with the same energy $\varepsilon_i$ and hybridization amplitude $\gamma_i$ within each group, by symmetry. The anomalous operators, $\Delta_i$, are represented by the dashed lines and can be modified to allow for different pairing symmetries.
	}
	\label{fig:schema_cdmft}
\end{center}
\end{figure}

To probe superconductivity using CDMFT, we need to include anomalous hybridizations between bath orbitals. In this work we use a 3-site cluster with 9 bath orbitals defined on Fig. \ref{fig:schema_cdmft}. The bath parametrization is based on the $C_{3v}$ symmetry. To control the type of superconductivity probed, we change the value of the coefficient $\chi$: Type $s$ corresponding to $\chi=1$ and type $d+id$ to $\chi = e^{\frac{2\pi i}{3}}$. Using this method we can probe one by one every pairing symmetry listed in Table \ref{table:pairing}.

%===============================================================================
\section{Electron doping}

\label{sec:electron}
\begin{figure*}[t!]
	\subfigure{
		\label{fig:supra_ep30}
		\includegraphics[scale=0.49]{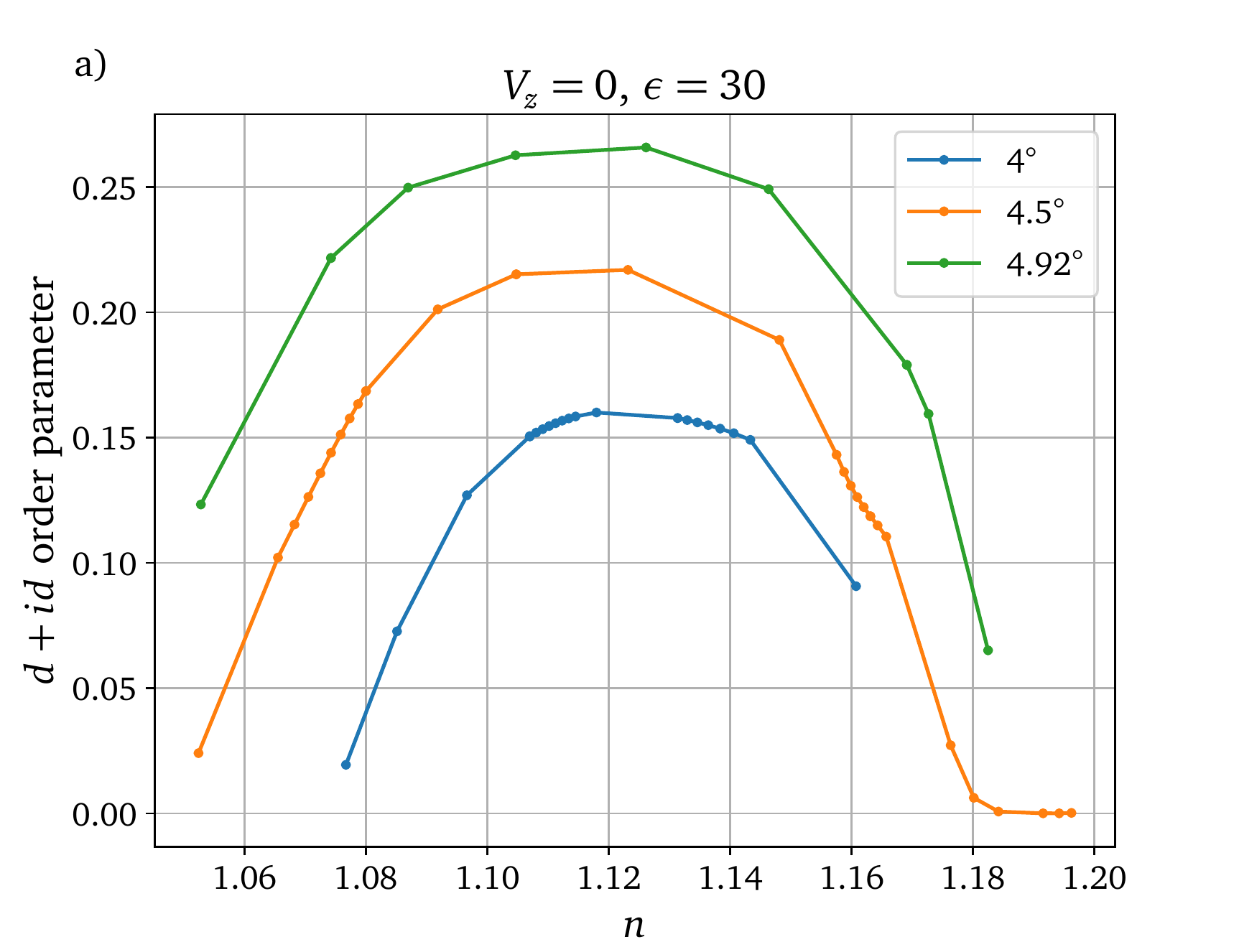}	}
	\quad
	\subfigure{
		\includegraphics[scale=0.49]{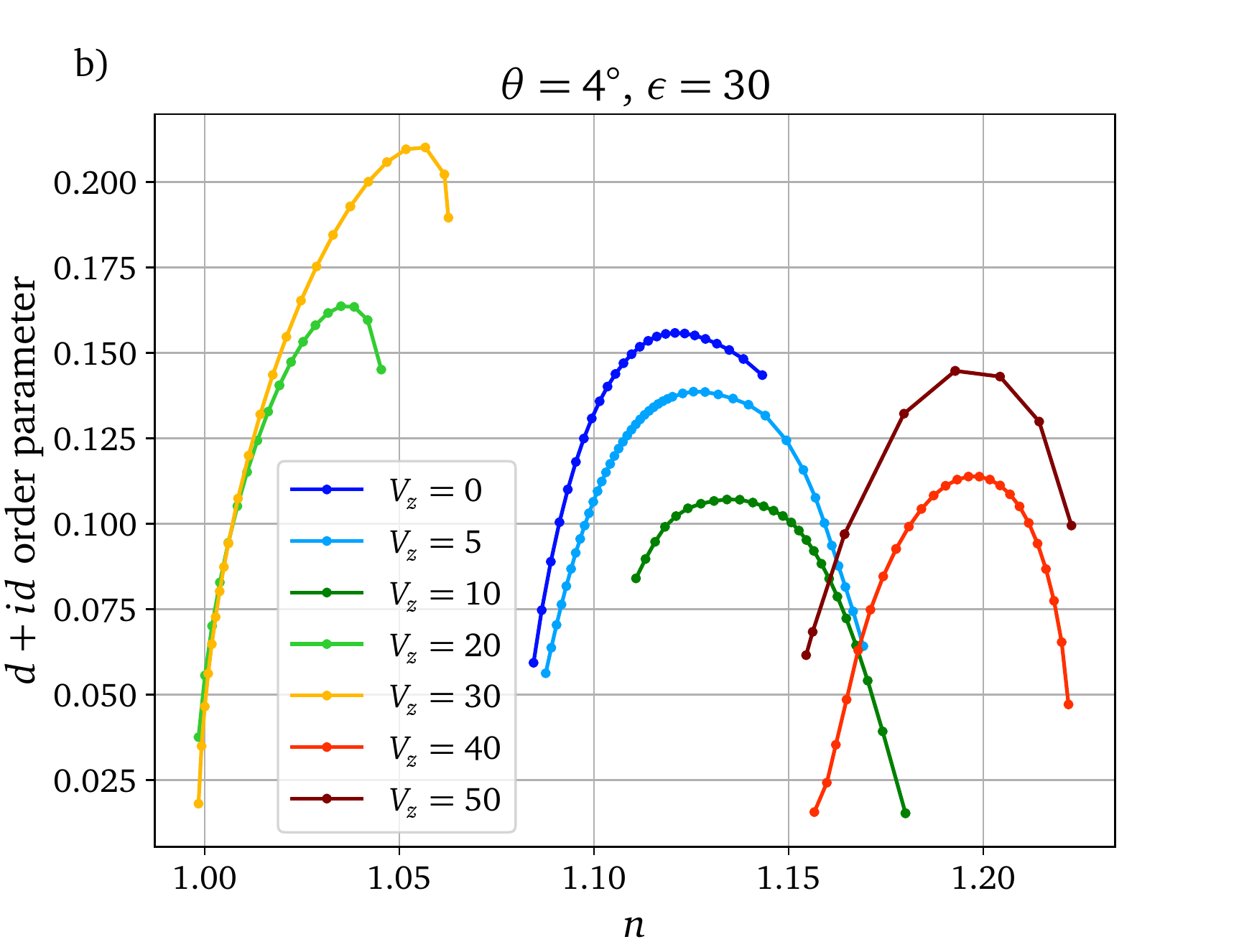}
		\label{fig:supra_Vz}}
	\caption{Amplitude of the $d+id$ order parameter on the electron doped-side obtained in the VCA for $\epsilon=30$. a) Effect of the twist angle on the order parameter function of density $n$. b) Effect of the interlayer potential $V_z$ (in meV) on the order parameter for $\theta=4^{\circ}$.}
\end{figure*}
We first start by the electron-doped side of the phase diagram. We use the VCA in this region in order to understand the effect of the twist angle $\theta$ and $V_z$. In all our VCA computations we used the 12-site cluster of Fig. \ref{fig:schema_model} with the Weiss field defined in Table \ref{table:pairing}.

We first look at the effect of the twist angle on the superconducting state in the absence of displacement field ($V_z=0$). We consider a typical value of the effective dielectric constant ($\epsilon=30$). The data from Ref. \cite{pan_band_2020}, gives us the parameters presented in Table \ref{table:hoppingVz0}, for the different twist angles considered. At $V_z=0$, the hopping parameters all have the same phase $\phi^s_n=-\pi$, which implies real-valued parameters.

%~~~~~~~~~~~~~~~~~~~~~~~~~~~~~~~~~~~~~~~~~~~~~~~~~~~~~~~~~~~~~~~~~~~~~~~~~~~~~~	

\begin{table}[h]

	\begin{ruledtabular}
		
\begin{tabular}{ccccc}
	$\theta$&$t_1$&$t_2$&$t_3$&$U$\\
	\hline
	$4^\circ$&$-1$&$-0.053$&$-0.116$&$10.61$\\

	$4.5^\circ$&$-1$&$-0.045$&$-0.135$&$7.78$\\

	$4.92^\circ$&$-1$&$-0.037$&$-0.149$&$6.24$\\	
\end{tabular}
\end{ruledtabular}
	\caption{Hopping parameters for the different twist angles $\theta$ considered for $V_z=0$ and $\epsilon=30$. In this case, all hopping parameters are real.\label{table:hoppingVz0}}
\end{table}
%~~~~~~~~~~~~~~~~~~~~~~~~~~~~~~~~~~~~~~~~~~~~~~~~~~~~~~~~~~~~~~~~~~~~~~~~~~~~~~	

We probed all the pairing symmetries enumerated in Table \ref{table:pairing} (the Weiss field is defined in the rightmost column). We found a non-trivial VCA solution with $d+id$ pairing, but none for the other pairing operators listed in Table \ref{table:pairing}, except $d-id$, which is degenerate with $d+id$.

In Fig. \ref{fig:supra_ep30}, we show the $d+id$ order parameter for $V_z=0$. A solution exists for all twist angles considered, for some interval of density $n$ ($n=1$ corresponding to half-filling). The order parameter is computed from the VCA Green function using Eq. (\ref{eq:expectation}).  We also observe that the superconducting dome is greatly affected by the twist angle. Increasing the twist angle widens the range of density where superconductivity exists. The maximum order parameter also increases with twist angle. This is caused by the drop of $U$ as $\theta$ increases. Assuming a monotonous relation between the order parameter and the critical temperature $T_c$, we can infer that, in the range of angles considered, $T_c$ can be tuned by changing the twist angle. Another reason $T_c$ depends on the twist angle $\theta$ is that our energy scale is defined by $t_1$, which itself increases as a function of $\theta$ when expressed in meV \cite{pan_band_2020} (the relation between $T_c$ and the order parameter depends on this energy scale). 

We then consider the effect of the external perpendicular displacement field. In Fig. \ref{fig:supra_Vz} we show the $d+id$ order parameter obtained from VCA for a twist angle $\theta=4^{\circ}$, for different values of $V_z$. Again, a nontrivial solution was found only for the $d\pm id$ order parameter. The doping range is greatly affected by the displacement field. It is noted in Ref. \cite{pan_band_2020} that the position of the van Hove singularity (VHS) depends on $V_z$ and crosses over the half-filling point at around $V_z=28$ meV. We expect that one of the factors affecting the doping range is this displacement of the VHS. This could explain why the maximum of the dome is closer to half-filling for values of $V_z$ close to $28$ meV. It could also explain the localization of the superconducting dome for higher values of $V_z$, when the VHS is beyond half-filling. As noted in Ref. \cite{klebl_competition_2022}, the instability occurs mostly around the VHS due to the high density of state. Our result seems to be in agreement with this.

We also used CDMFT in this region, but the procedure did not converge for every twist angle and value of $V_z$.

%===============================================================================
\section{Hole doping}
\label{sec:hole}

\begin{figure*}[t!]
	\subfigure{
		\label{fig:supra_cdmft}
		\includegraphics[scale=0.49]{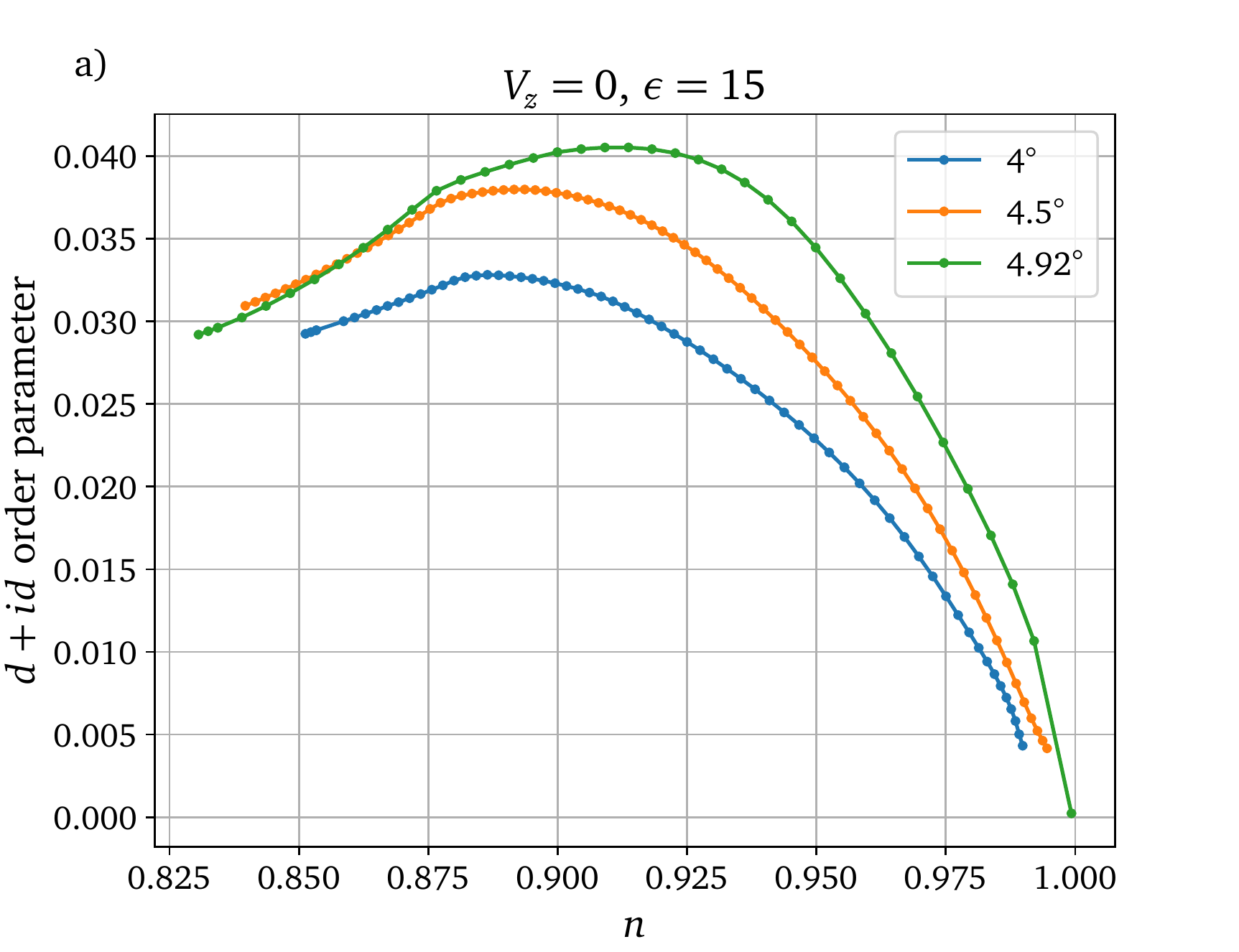}	}
	\quad
	\subfigure{
		\includegraphics[scale=0.49]{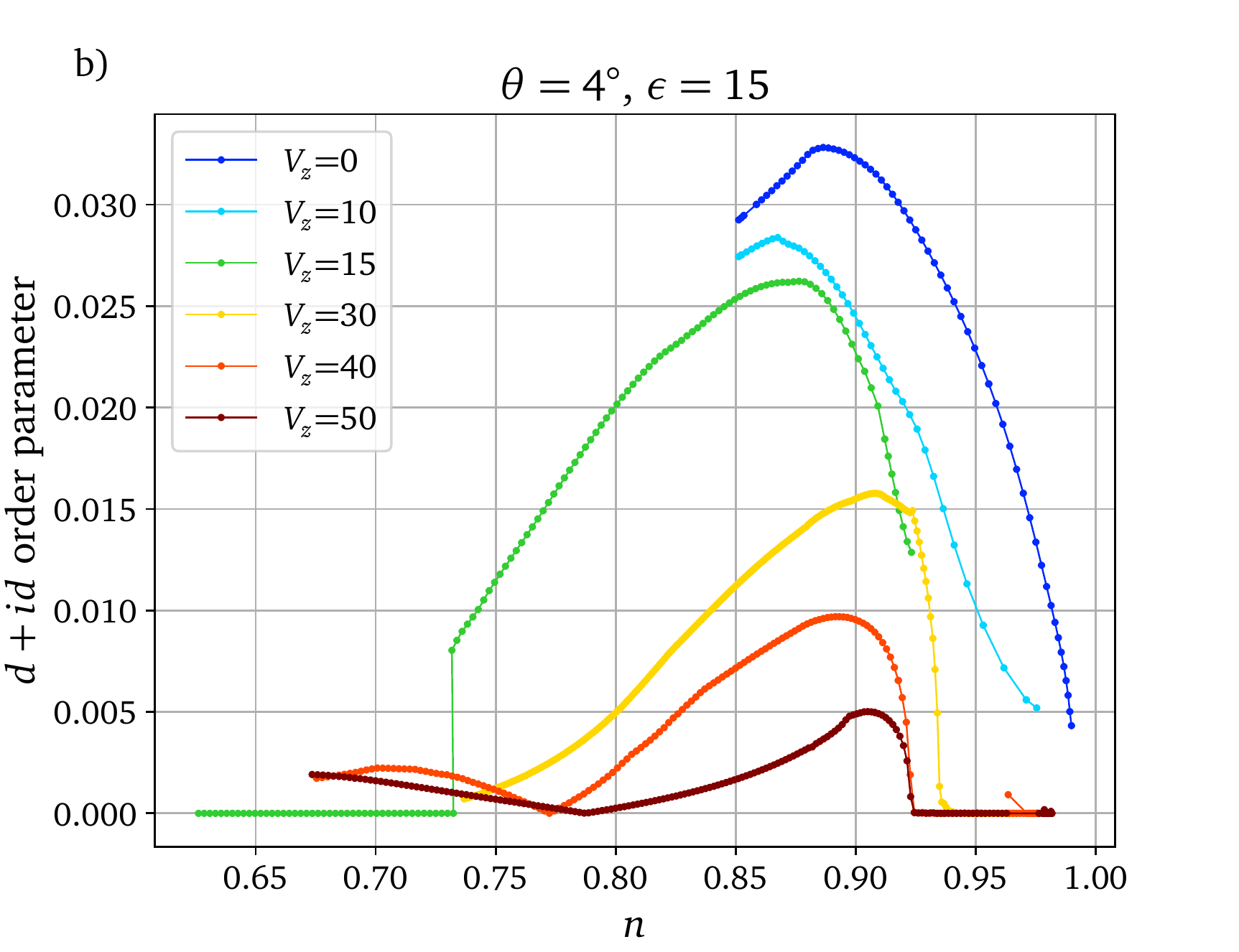}
		\label{fig:supra_VZ_cdmft}}
	\caption{Amplitude of the $d+id$ order parameter on the hole doped-side obtained from CDMFT for $\epsilon=15$. a) Effect of the twist angle on the order parameter function of density $n$. b) Effect of the interlayer potential $V_z$ (in meV) on the order parameter for $\theta=4^{\circ}$.}
\end{figure*}

We did not find any VCA solution on the hole-doped side. This was caused by a discontinuity in the Potthoff self-energy functional (\ref{eq:potthoff}) as a function of the Weiss field, leading to an invalid solution of Eq. (\ref{eq:min}) or no solution at all. This reflects the structure of the energy levels on the cluster : on the hole-doped side, the energy levels seem to be more closely spaced than on the electron-doped side, which may lead to those discontinuities. This does not mean that superconductivity is not present. In fact, by using CDMFT we are able to extract information on this region. To do so, we use the method and cluster from section \ref{sec:cdmft}. Note that the value of $U$ in the Hubbard model that gives access to the Mott physics depends on cluster size \cite{dang_mott_2015}. To make sure we are considering similar physical properties with a 3-site cluster, we set the dielectric constant $\epsilon$ to $15$, which doubles the value of $U$. As mentioned before, $\epsilon$ is determined by the sample's environment and $\epsilon=15$ is still in the range of values seen in the literature.

In Fig. \ref{fig:supra_cdmft}, we show the $d+id$ order parameter as a function of density, obtained from the CDMFT procedure at $V_z=0$. We observe that the maximum of the order parameter is affected by the twist angle. The increase in the order parameter with twist angle follows what was observed on the electron-doped side. The main difference here is that the width and position of the dome seem to be less affected by the twist angle. The presence of a $d+id$ superconducting phase on the hole-doped side agrees with predictions about the triangular-lattice $t$-$J$ model \cite{huang_emergent_2022}.

For this set of parameters, we do not observe other types of superconductivity. We probed the other pairing symmetries by changing the value of $\chi$ in Fig. \ref{fig:schema_cdmft} (e.g. from $e^{\frac{2\pi i}{3}}$ to $1$ to probe extended s-wave pairing) and by changing the type of nearest-neighbor anomalous operator from $S_{\mathbf{r},i}$ to $T_{\mathbf{r},i}$  (singlet to triplet). In all cases, the order parameter vanishes for the range of doping accessible. 

On figure \ref{fig:supra_VZ_cdmft}, we show the $d+id$ order parameter for different values of $V_z$ and a fixed twist angle of $4^{\circ}$. The effect of the displacement field is different from what was observed on the electron-doped side using VCA. The superconducting dome is greatly affected by the modulation of $V_z$. As observed in the figure, the maximum of the parameter decreases with $V_z$. We also observe that the superconducting dome is suppressed around half-filling when $V_z$ is increased. Here again, we did not find any solution for the other pairing symmetries.

%===============================================================================
\section{Discussion}
\label{sec:discussion}

Our results agree with the evidence of superconductivity presented in Ref. \cite{wang_correlated_2020}, were a zero resistivity state was observed when doping away from half-filling for a twist angle of $5.1^\circ$, in the presence of a displacement field. The two superconducting domes are separated by an insulating state around half-filling. We predict a superconducting phase on both electron- and hole-doped sides. We also predict that the two superconducting domes are separated from each other by a finite range of densities that includes an insulating state.

While looking at the effect of the twist angle, the order parameter seems to be mostly affected by the on-site interaction $U$. The shrinking of the superconducting dome seems to be related to increasing the interaction $U$. It has been shown before for the square-lattice Hubbard model, that the maximum of the superconducting dome decreases with $U$ \cite{senechal_competition_2005,dash_pseudogap_2019}. The values of $U$ considered in this work are in the strong interaction range. The $U$ dependence observed here agrees with the expected behavior. 

The effect of $V_z$, is different on both sides of half-filling. As stated above, we expect that the behavior on the electron-doped side is linked to the shift in the VHS. However, we do not think the same explanation works on the hole-doped side, since the maximum of the dome does not shift according to the VHS. This difference might be explained by the choice of numerical method and by the difference in the cluster used.

The effect of the displacement field on the order parameter in the hole-doped region does not seem to come from the change in the interaction $U$. In fact, while we increase $V_z$, $U$ decreases but, contrary to what is expected, the superconducting dome also decreases, even though we are still in the strong interaction regime. This is likely related to the complex hopping parameter induced by $V_z$ : A phase factor is introduced on the bonds, which produces a non-zero staggered flux on each plaquette. In our CDMFT calculation, only one plaquette is considered, which implies a non-zero total flux. We expect that this flux, the equivalent of a magnetic field, pushes the superconducting phase down. To explore this, we tried a 4-site system with 8 baths to have two plaquettes with total flux of zero. Unfortunately, due to the loss of the $C_{3v}$ symmetry on that cluster, no meaningful results were found. This effect seem to be less present in the VCA procedure since the 12-site cluster is more balanced even if the total flux is still non-zero. It would be interesting to isolate the contribution of the complex phases and observe its effect on the superconducting dome. 

In both methods used here we found a $d\pm id$ solution that was greatly affected by the perpendicular displacement field. We explored the possibility of other pairings, like those studied in Ref. \cite{wu_pair-density-wave_2022}, but did not detect them. This can be explained by the presence of second- and third-nearest neighbor hopping terms in Model (\ref{eq:H}), and by the numerical methods used here that allow us to probe larger values of the interaction $U$.

In both VCA and CDMFT, we tried different clusters with varying sizes and symmetries. To have a proper convergence of the algorithms, the cluster needs to respect a set of criteria. Like stated above, the most important condition is to respect the $C_{3v}$ group symmetry. This allows us to separate the different order parameters in each irreducible representation. This also provides stability to the procedure. Moreover, the use of exact diagonalization as impurity solver limits the total number of orbitals that can be realistically considered to 12, because of the computational resources needed. We could not used a quantum Monte Carlo impurity solver; indeed, the fermion sign problem is strongly enhanced by the triangular lattice and, in any case, complex order parameters cannot be studied with that method. This makes exact diagonalization the only practical impurity solver in this case, and the limited cluster sizes precludes any finite-size analysis. We also found a better convergence around low doping when the total number of orbitals was even. For example, a cluster of 7 sites gave rise to a discontinuity in the Potthoff functional close to half-filling, which translated into convergence issues in the VCA algorithm. This was likely due to the unpaired electron that shifted the cluster's energy levels. All those subtleties in the numerical procedure made it harder to probe the effect of cluster size and shape.

%===============================================================================
\section{Conclusion}

We have applied quantum cluster methods to an effective low-energy model for a twisted bilayer of WSe${}_2$, developed in Refs \cite{jung_ab_2014,pan_band_2020}, and found a superconducting dome in both electron- and hole-doped regions in the strong-interaction regime. In both cluster methods used, VCA and CDMFT, we only found superconductivity with $d\pm id$ pairing symmetry. The twist angle can be used to modify the maximum order parameter. If a perpendicular displacement field is applied, the superconducting dome is modified greatly because of the complex phase acquired by the hopping parameters. All those predictions are in agreement with previous experimental evidence \cite{wang_correlated_2020} of superconductivity in twisted bilayer WSe${}_2$.

\begin{acknowledgments}
This work has been supported by the Natural Sciences and Engineering Research Council of Canada (NSERC) under grant RGPIN-2020-05060 and by NSERC postgraduate scholarships doctoral program.
Computational resources were provided by Digital Research Alliance of Canada and Calcul Qu\'ebec.
\end{acknowledgments}

%===============================================================================
%\bibliography{biblio}  % d\'ecommenter si biblio.bib existe
%

%\appendix

%\section{Parameter used in the Hubbard model}

\end{document}